\documentstyle[12pt,epsf]{article}

\begin{document}
\begin{flushright}
UB/FI 98-2\\
September 1998\\
To be published in Physics Letters B
\end{flushright}

{\LARGE \bf
\begin{center}
Three-Flavour Neutrino-Mixing Implications
of the LSND Result
\end{center}
}

\vspace{0.2em}

\begin{small}
\begin{sloppypar}
\noindent
\begin{center}
G.~Conforto{~\footnote {\tt <conforto@cern.ch>}}\\
CERN, European Laboratory for Particle Physics \\
CH-1211 Gen\`eve 23, Switzerland \\
\vskip .03in
and\\
\vskip .03in
M.~Barone{~\footnote {\tt <maura@fis.uniurb.it>}} and
C.~Grimani{~\footnote {\tt <cgrimani@fis.uniurb.it>}}\\
Universit\`a degli Studi, I-61029 Urbino, Italy
\end{center}
\end{sloppypar}
\end{small}

\begin{abstract}
 The LSND result is shown to fit into a minimal three-flavour
 neutrino-mixing scenario capable of describing all known experimental
 facts provided the large $\Delta M^2 = m_3^2 - m_2^2 \sim m_3^2 - m_1^2$
 lies in the range $2.5 \times 10^{-1} < \Delta M^2 < 3.0~{\rm eV}^2$. In
 this range the value of $P_{\mu\tau}$ is expected to be about 5\% or
 larger.
\end{abstract}

\section{Introduction}

In recent years, several unexpected results have appeared in accelerator,
atmospheric and solar neutrino physics~\cite{conrad}. Although none of them
is really beyond questioning, collectively they represent a fair evidence
for the existence of some new phenomenon.

Taken singularly, all these results can be explained in terms of neutrino
oscillations~\cite{pontecorvo}. However, this interpretation can only be
viable if all existing positive and negative results can be accounted for
by a unique set of oscillation parameters.

In this paper we examine whether there exist some conditions under which
the LSND result~\cite{white}, however controversial it may
be~\cite{karmen}, may fit into a minimal three-flavour neutrino-mixing
scenario constrained by all experimental observations. We show that such a
global description is possible. Ranges of values of the oscillation
parameters for which this occurs are given.

\section{Three-flavour neutrino-mixing phenomenology}

In the complete three-flavour approach, the weak eigenstates $|\nu_\alpha
\rangle$ = $\nu_e$, $\nu_\mu$, $\nu_\tau$ and the mass eigenstates $|\nu_i
\rangle$ = $\nu_1$, $\nu_2$, $\nu_3$ are related by a unitary
transformation matrix U. The probability of an initial neutrino
$\nu_\alpha$ of energy E being equal to another neutrino $\nu_\beta$ at a
distance $L$, can be written as
\begin{eqnarray}
    P_{\alpha \beta} = 
\delta_{\alpha \beta} - 4 \sum_{j>i} U_{\alpha i} U_{\beta i} U_{\alpha j}
U_{\beta j} \sin^{2}(\Delta_{ij}/2)              \label{eq:Pab}
\end{eqnarray}
with $\Delta_{ij} = \Delta m_{ij}^2 L/2E$ , where
$\Delta m_{ij}^2 = m_i^2 - m_j^2, m_i = m_{\nu_{i}}$.  

Assuming $CP$-invariance, the U-matrix is real and can be parametrized as
\[
\left(
\begin{array}{ccc}
c_{12}c_{13}                      
&s_{12}c_{13}                       &s_{13}     \\
-s_{12}c_{23} - c_{12}s_{23}s_{13}
&c_{12}c_{23} - s_{12}s_{23}s_{13}  &s_{23}c_{13}  \\      
s_{12}s_{23} - c_{12}c_{23}s_{13} 
&-c_{12}s_{23} - s_{12}c_{23} s_{13}&c_{23}c_{13} \\
\end{array}
\right)
\]
with $c_{ij} = \cos \theta_{ij}$ and $s_{ij} = \sin \theta_{ij}$, where
$\theta_{12}$, $\theta_{13}$ and $\theta_{23}$ are three independent real
angles lying in the first quadrant.

Of the three $\Delta m_{ij}^2$'s appearing in eq.~(\ref{eq:Pab}), only two
are independent. Therefore, the complete solution of the problem consists
in determining five unknowns: two $\Delta m_{ij}^2$'s and the three
$\theta_{ij}$'s.

Under the additional hypothesis of a natural mass hierarchy $m_1 << m_2 <<
m_3$, the oscillatory behaviour of eq.~(\ref{eq:Pab}) is determined by the
large $\Delta M^2 = m_3^2 - m_2^2 \sim m_3^2 - m_1^2$ and the small $\Delta
m^2 = m_2^2 - m_1^2$. The transition probabilities are then given by the
sum of two terms describing, respectively, the fast oscillation
(characterised by $\Delta M^2$) and the slow oscillation (characterised by
$\Delta m^2$).

Lastly, the assumption of the dominance of diagonal terms in the mixing
matrix, implying a strong correlation between flavour and mass eigenstates,
ensures that $s_{ij} < c_{ij}$. The angles $\theta_{ij}$'s in the mixing
matrix are then uniquely defined.

\section{Experimental inputs and results}

The oscillation analysis of the atmospheric neutrino data implies a $\delta
m^2$ of about $10^{-3}~{\rm eV}^2$~\cite{kajita,superk}. Such a small
$\delta m^2$ cannot account for the LSND observation of
$(P_{e\mu})_{\rm LSND} = (3.1 \pm 0.09 \pm 0.05) \times 10^{-3}$ at $L/E
= 0.7$~m/MeV. Consequently, it must be identified with the smaller $\Delta
m^2$, thus implying $\Delta m^2 \sim 10^{-3}~{\rm eV}^2$.

This defines the two ranges $L/E < 10^3$~m/MeV, in which the transition
probabilities $(P_{\alpha \beta})_1$ are dominated by the fast component,
and $L/E > 10^3$~m/MeV in which the transition probabilities $(P_{\alpha
\beta})_2$ depend on both the slow and fast components. The average
transition probabilities $\langle P_{\alpha \beta}\rangle_1$ and $\langle
P_{\alpha \beta}\rangle_2$ are calculated from eq.~(\ref{eq:Pab}) for
$\sin^2(\Delta_{12}/2)=0$, $\langle \sin^2(\Delta_{13}/2)\rangle =0.5$ and
$\langle \sin^2(\Delta_{12}/2)\rangle=0.5$, $\langle
\sin^2(\Delta_{13}/2)\rangle =0.5$, respectively.

$\Delta m^2 \sim 10^{-3}~{\rm eV}^2$ implies in turn the
energy-independence of all oscillation phenomena occurring in solar
neutrinos. This is consistent with the present experimental situation.

The energy-independence of the solar neutrino deficit has been long
advocated~\cite{conforto1,perkins,acker,conforto2,conforto3}. Using the
most recent experimental results~\cite{lande,gavrin,kirsten,suzuki} and
theoretical predictions~\cite{bahcall} but neglecting solar model
systematic errors~\cite{conforto3,bahcall}, a fit for an energy-independent
oscillation-induced depletion of the $\nu_e$ flux yields the result
\begin{eqnarray}
                \langle P_{ee}\rangle_2 = 0.50 \pm 0.06      \label{eq:Pee}
\end{eqnarray}
with a confidence level C.~L. = 0.34 \%. This is admittedly marginal but,
in view also of the many sometimes optimistic approximations, not
unacceptably small.

The day- and night-spectra measured by the Super-Kamiokande experiment
also show no anomalous behaviour. The fit for the same energy-independent
suppression of the $\beta$-decay expectations in both spectra yields a
confidence level C.~L. = 1.6\%~\cite{suzuki}.

The over-all confidence level including all the information from solar
neutrino rates, day/night effect and energy spectrum shape is C.~L. =
0.26\%, largely dominated by the marginal consistency among rates. Thus,
even neglecting all caveats~\cite{conforto2}, a deviation from
energy-independence in solar neutrinos has at the most the significance of
a $3\sigma$ effect.

A second consequence of $\Delta m^2 \sim 10^{-3}~{\rm eV}^2$ is that for
$L/E < 10^3$~m/MeV all oscillation phenomena depend only on the two angles
$\theta_{13}$ and $\theta_{23}$, the influence of the third angle
$\theta_{12}$ becoming sizeable only for $L/E = 10^3$~m/MeV or greater.

Two-flavour analyses of transition probability results are normally
presented in terms of contours in the $\sin^2(2\theta)$, $\delta m^2$
plane. From eq.~(\ref{eq:Pab}) it is easy to see that, for $L/E <
10^3$~m/MeV, the $\sin^2(2\theta)$'s relative to the transitions
$\nu_e$ - $\nu_x$, $\nu_e$ - $\nu_\mu$ and $\nu_\mu$ - $\nu_\tau$ are in
fact, respectively, the three-flavour oscillation amplitudes $(A_{ex})_1 =
4 s_{13}^2 c_{13}^2$, $(A_{e\mu})_1 = 4s_{23}^2 s_{13}^2 c_{13}^2$ and
$(A_{\mu\tau})_1 = 4s_{23}^2 c_{23}^2 c_{13}^4$ .

Thus, from the knowledge of the maximum values experimentally allowed for
the first and the third of these amplitudes as a function of $\Delta
M^2$~\cite{nomad,achkar,apollonio}, an upper limit contour for the second
can be readily determined.

\begin{figure}
\begin{center}
\leavevmode
\epsfxsize \textwidth
\epsffile{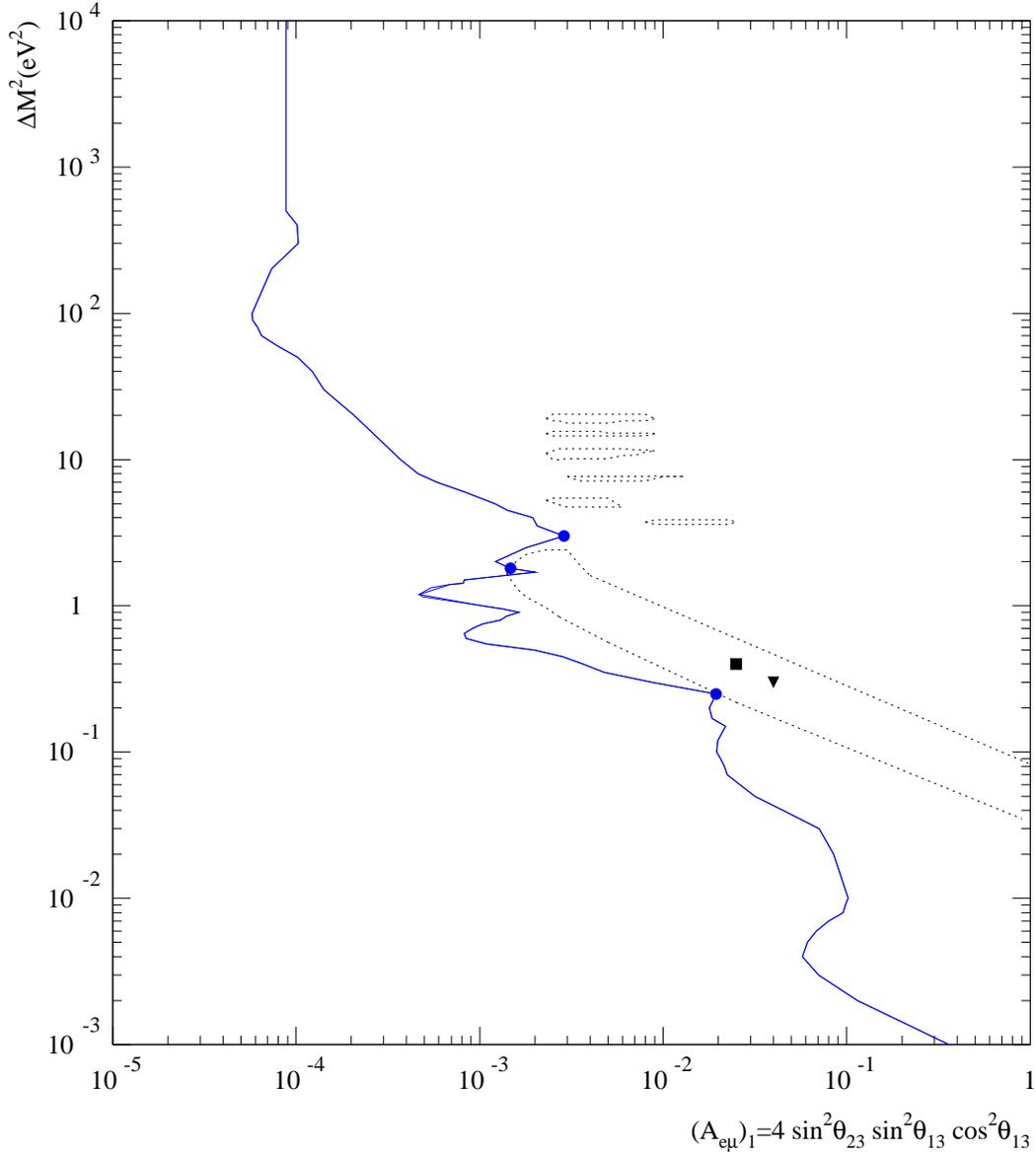}
\end{center}
\caption[]{The upper limit on the oscillation amplitude $(A_{e\mu})_1=4
s_{23}^2 s_{13}^2 c_{13}^2$ calculated from the maximum values
experimentally allowed for $(A_{ex})_1 = 4 s_{13}^2 c_{13}^2$ and
$(A_{\mu\tau})_1=4s_{23}^2 c_{23}^2 c_{13}^4$ as a function of $\Delta M
^{2}$ (full line) compared with the 99\% C.~L. region allowed by the LSND
result (dotted lines) . The dots correspond to the choices of the angles
$\theta_{23}$ e $\theta_{13}$ for three typical values of $\Delta M^2$.
The square and the triangle represent two solutions recently proposed.}
\label{fig:1}
\end{figure}

The limit-curve for $(A_{e\mu})_1 = 4 s_{23}^2 s_{13}^2 c_{13}^2$ as a
function of $\Delta M^2$ is shown in Figure~\ref{fig:1} together with the
99\% C.~L. LSND-allowed region. Their compatibility clearly restricts
$\Delta M^2$ approximately to the range
\[
2.5 \times 10^{-1} < \Delta M^2 < 3.0~{\rm eV}^2.
\]

Within this range, three different situations may occur in an experiment
centered around a typical $L/E$ of $0.7$~m/MeV. The term $\sin^{2}(\Delta
M^2 L/4E)$ is close to the average value of 0.5 at the upper end of the
range, reaches its maximum in the central region and falls to small values
at the lower end.

The angles $\theta_{13}$ and $\theta_{23}$ obviously depend on $\Delta M^2$
and on the amount by which the limit of Figure~\ref{fig:1} is accepted to
be violated in order to reach the LSND-allowed region. If, in order to
avoid exceeding any limit, they are conservatively chosen to coincide with
their upper limits, for the three typical choices of $\Delta M^2$ above
they take the values reported in Table~\ref{tab:small}. These solutions are
indicated by dots in Figure~\ref{fig:1}.

\begin{table}
\begin{center}
\def\arraystretch{1.2}
\begin{tabular}{|c|c|c|c|}
\hline
\hline
$\Delta M^2$ (${\rm eV}^2$)
& 3.0                & 1.8                  & 0.25    \\
\hline
\hline
$s_{13}^2$                 
&$3.6\times 10^{-2}$ &$2.0\times 10^{-2}$   &$9.8\times 10^{-3}$ \\
\hline
$s_{23}^2$                 
&$2.0\times 10^{-2}$ &$1.8\times 10^{-2}$   & 0.50   \\
\hline
$(P_{e\mu})_{\rm LSND}$    
&$1.4\times 10^{-3}$ &$1.4\times 10^{-3}$   &$0.94\times 10^{-3}$ \\
\hline
$(P_{\mu\tau})_{\rm LSND}$ 
&$3.6\times 10^{-2}$ &$6.8\times 10^{-2}$   &$4.8\times 10^{-2}$ \\
\hline
$\langle P_{ee}\rangle_1$               
&0.93                & 0.96                 &0.98  \\
\hline
$\langle P_{e\mu}\rangle_1$             
&$1.4\times 10^{-3}$ &$0.71\times 10^{-3}$  &$9.7\times 10^{-3}$ \\
\hline
$\langle P_{\mu\mu}\rangle_1$           
&0.96                &0.97                  &0.50 \\
\hline
$\langle P_{\mu\tau}\rangle_1$          
&$3.6\times 10^{-2}$ &$3.4\times 10^{-2}$   &0.49 \\
\hline
$r(\Theta={\rm small})$    
&1.0                 &1.0                   &0.50\\
\hline
\hline
\end{tabular}
\end{center}
\caption[]{
Mixing angles and transition probabilities for the three types of solutions
allowed by the LSND result for $L/E < 10^3$~m/MeV.

$(P_{\alpha \beta})_{\rm LSND}$ represents a transition probability
calculated for the LSND value of $L/E=0.7$~m/MeV,
$\langle P_{\alpha \beta}\rangle_1$ is
the same quantity averaged after the onset of the fast oscillation,
$r(\Theta={\rm small})$ is the ratio between the observed and expected
$N_\mu/N_e$ ratio measured in the Superkamiokande experiment at short $L$.}
\label{tab:small}
\end{table}

In the allowed mass range above, $s_{13}^2$ is constrained by the tight
reactor limits and cannot have large variations whilst $s_{23}^2$ can swing
by as much as a factor of twentyfive. However, the transition rates at the
LSND value $L/E=0.7$~m/MeV are relatively constant around the low-side
values $(P_{e\mu})_{\rm LSND} =1 \times 10^{-3}$ ($2\sigma$ down
relative to the LSND result) and $(P_{\mu\tau})_{\rm LSND}
=5\times10^{-2}$.

The Superkamiokande experiment has measured the ratio $r$ between the
observed and expected $\nu_\mu$/$\nu_e$ ratios in cosmic rays as a function
of the zenith angle $\Theta$~\cite{kajita,superk}. Small values of $\Theta$
correspond to $L/E < 10^3$~m/MeV so that $r(\Theta={\rm small})$ can be
calculated from the the average rates $\langle P_{\alpha \beta}\rangle_1$
reported in Table~\ref{tab:small} through the relation $r = (P_{\mu\mu} +
\rho P_{e\mu})/(P_{e\mu}/\rho + P_{ee})$ where $\rho = 0.47 \pm 0.02$ is
the expected $\nu_e/\nu_\mu$ flux ratio in the absence of
oscillations~\cite{conforto1}. Depending on $\Delta M^2$,
$r(\Theta={\rm small})$ varies between 0.5 and 1.0. The Superkamiokande
data indicate a value around 0.8. More precise data could help restricting
the range of allowed parameters around the LSND result.

Two most recently suggested solutions $\Delta M^2 = 0.4~{\rm eV}^2$,
$s_{13}^2= 3.2\times10^{-2}$, $s_{23}^2=0.2$~\cite{thun} and $\Delta M^2 =
0.3~{\rm eV}^2$, $s_{13}^2=5.1\times 10^{-2}$,
$s_{23}^2=0.21$~\cite{barenboim} are also indicated in Figure~\ref{fig:1}
by a square and a triangle, respectively.

The third angle $\theta_{12}$ can easily be determined from
eq.~(\ref{eq:Pee}).
This yields the results reported in Table~\ref{tab:large}.

\begin{table}
\begin{center}
\def\arraystretch{1.2}
\begin{tabular}{|c|c|c|c|}
\hline
\hline
$\Delta M^2$ (${\rm eV}^2$)
&3.0                 &1.8                 &0.25    \\
\hline
\hline
$s_{13}^2$              
&$3.6\times 10^{-2}$ &$2.0\times 10^{-2}$ &$9.8\times 10^{-3}$ \\
\hline
$s_{23}^2$              
&$2.0\times 10^{-2}$ &$1.8\times 10^{-2}$ &0.50 \\
\hline
$s_{12}^2$              
&0.36                &0.40                &0.43 \\
\hline
$\langle P_{e\mu}\rangle_2$          
&0.44                &0.47                &0.26 \\
\hline
$\langle P_{e\tau}\rangle_2$         
&0.05                &0.03                &0.24 \\
\hline
$\langle P_{\mu\mu}\rangle_2$        
&0.51                &0.50                &0.37 \\
\hline
$\langle P_{\mu\tau}\rangle_2$       
&0.05                &0.04                &0.37 \\
\hline
$r(\Theta={\rm large})$ 
&0.49                &0.48                &0.47 \\
\hline
\hline
\end{tabular}
\end{center}
\caption[]{
Mixing angles and transition probabilities for the three types of solutions
allowed by the LSND and solar neutrino results for $L/E > 10^3$~m/MeV.

$\langle P_{\alpha \beta}\rangle_2$ represents a transition probability
averaged after
the onset of the slow oscillation, $r(\Theta={\rm large})$ is the ratio
between the observed and expected $N_\mu/N_e$ ratio measured in the
Superkamiokande experiment at long $L$.}
\label{tab:large}
\end{table}

The salient feature in the region $L/E > 10^3$~m/MeV is the presence of a
large $\nu_e$ - $\nu_\mu$ transition. The value of $r$ for long $L$,
$r(\Theta={\rm large})$, is practically constant at a value slightly
below 0.5, in good agreement with the Superkamiokande data.

\begin{figure}
\begin{center}
\leavevmode
\epsfxsize \textwidth
\epsffile{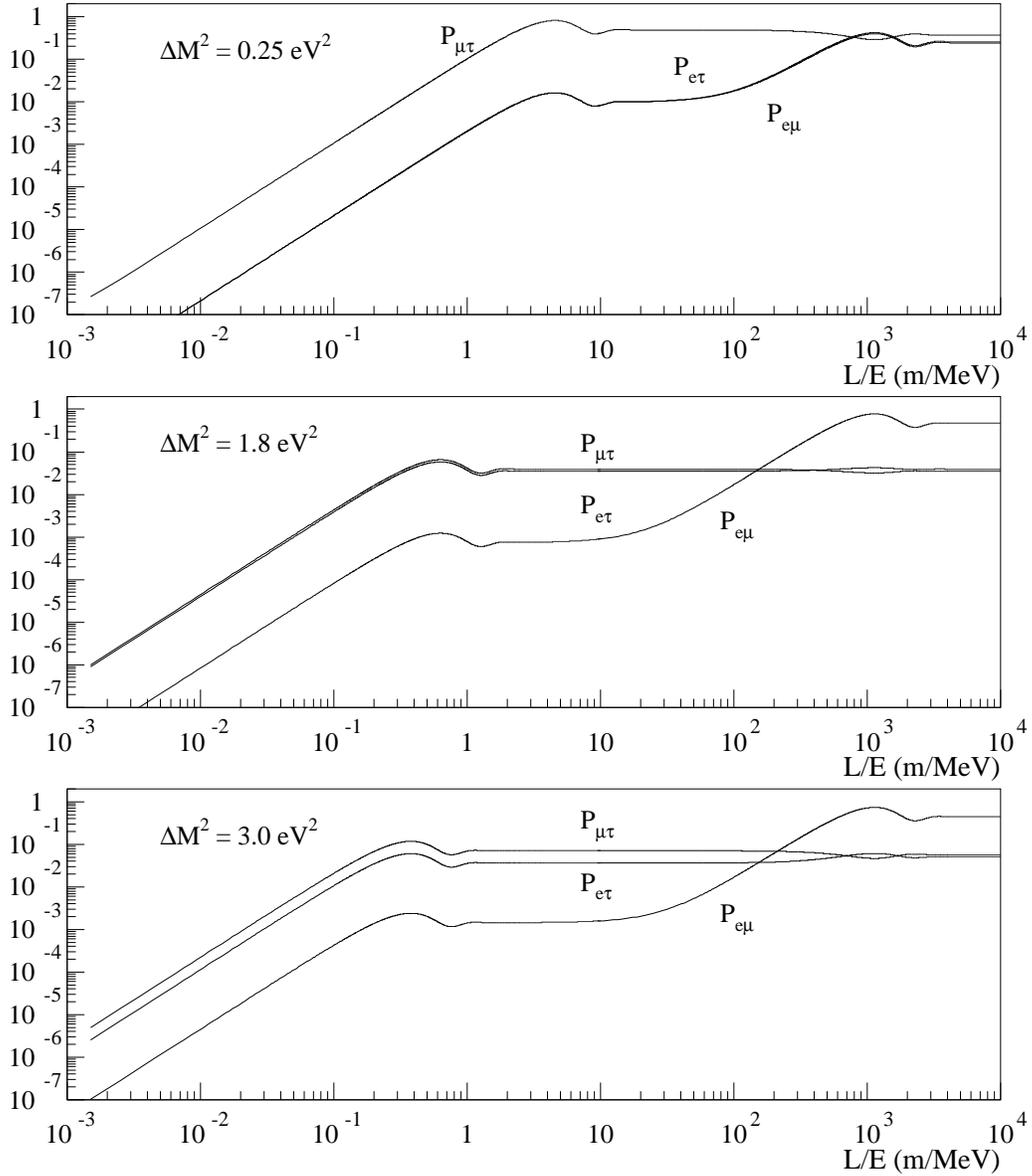}
\end{center}
\caption[]{The transition probabilities $P_{\alpha \beta}$
calculated as a function of $L/E$ for the three typical values of
$\Delta M^2$. The curves are averaged over a Gaussian $L/E$ distribution
with 30\% width.}
\label{fig:2}
\end{figure}

The behaviours of the $P_{\alpha \beta}$'s as a function of $L/E$ for
various values of $\Delta M^2$ are shown in Figure~\ref{fig:2}. The curves
are averaged over a Gaussian $L/E$ distribution with 30\% width.

For the smallest value of $\Delta M^2$ (0.25 ${\rm eV}^2)$, the
similarity of $P_{e\mu}$ and $P_{e\tau}$ implies the expectation of a
vanishing up/down $\nu_e$ asymmetry $A_{e}$ in the Superkamiokande
experiment, in complete agreement with the measured
value \cite{kajita,superk}.

On the other hand, for the two larger values of $\Delta M^2$ (1.8 and 3.0
${\rm eV}^2$), the dominance of $P_{e\mu}$ over $P_{e\tau}$ for $L/E >
10^3$~m/MeV corresponds to positive values of $A_{e}$. This is still an
open possibility as the over-all fit to the Superkamiokande data of the
two-flavour hypothesis of $\nu_e$ - $\nu_\mu$ oscillations alone is quite
acceptable (C.~L. = 4.4\%) and yields a large $P_{e\mu}$ for $L/E >
10^3$~m/MeV ($\sin^2(2\theta)=0.93$)~\cite{kajita,superk}.

Future better data may be able to clarify this issue and possibly further
restrict the acceptable range of $\Delta M^2$.

\section{Conclusions}

Standard three-flavour neutrino-mixing phenomenology, supplemented by the
hypotheses of natural mass hierarchy ($m_1 << m_2 << m_3$) and strong
correspondence between flavour and mass eigenstates ($s_{ij} < c_{ij}$), is
quite adequate to interpret all neutrino phenomena observed so far.

Together with the oscillation analysis of the atmospheric neutrino data,
the LSND result implies a $\Delta m^2$ of about $10^{-3}~{\rm eV}^2$.

For $L/E < 10^3$~m/MeV all phenomena depend only on the two angles
$\theta_{13}$ and $\theta_{23}$. The available upper limits on the
quantities 1 - ($P_{ee})_1$ and ($P_{\mu\tau})_1$ provide enough
information to calculate the upper limit on $(A_{e\mu})_1 = 4 s_{23}^2
s_{13}^2 c_{13}^2$ as a function of $\Delta M^2$ shown in
Figure~\ref{fig:1}. From the compatibility with the LSND result, the
allowed range of $\Delta M^2$ is
\[
2.5 \times 10^{-1} < \Delta M^2 < 3.0~{\rm eV}^2,
\]
implying the immediate cosmological consequence that the Universe cannot be
closed by neutrinos alone.

Experiments studying the range $L/E \sim 1$~m/MeV~\cite{camilleri,joe} have
clearly optimal chances to detect the characteristic feature of
oscillations, namely a modulation as a function of $L/E$. On the basis of
the observed $\nu_\mu$ - $\nu_e$ transition probability, conservatively
taken as $(P_{e\mu})_{\rm LSND} \sim 1\times 10^{-3}$, a
$(P_{\mu\tau})_{\rm LSND} \sim 5\times 10^{-2}$ is expected.

It should be noted that $(P_{e\mu})_1$ and $(P_{\mu\tau})_1$ are linked by
the relation
\[
(P_{e\mu})_1 / (P_{\mu\tau})_1 = s_{13}^2/(c_{13}^2 c_{23}^2)
\]
so that an increase in $(P_{e\mu})_1$ implies necessarily a larger
$(P_{\mu\tau})_1$.

The calculated values of $r$ are in good agreement with the so far not very
accurate measurements reported by the Kamiokande experiment. However, the
comparison between calculations and experimental results is really
meaningful only in the two extremes of the $L/E$ range, the region
in-between depending on many experimental features. It should be emphasized
that a value of $r(\Theta={\rm small})$ measured to be significantly
below 1, together with the smallness of $(P_{e\mu})_{\rm LSND}$ and of
$1-\langle P_{ee}\rangle_1$ would constitute an experimental evidence for a
sizeable $\langle P_{\mu\tau}\rangle_1$.

The large $\nu_e$ - $\nu_\mu$ transition expected for $L/E > 10^3$~m/MeV is
compatible with the results of the CHOOZ experiment~\cite{apollonio} only
for $\Delta m^2 \sim 10^{-3}~{\rm eV}^2$ or smaller.

\section{Acknowledgements}

Many useful discussions with D.~Autiero, B.~Cousins, U.~Dore, L.~di~Lella,
A.~Marchionni, P.~Nedelec, H.~Pessard and F.~Vannucci are gratefully
acknowledged.

\end{document}